\pdfoutput=1
\documentclass[11pt,twoside]{article}
\usepackage{./asp2014}

\aspSuppressVolSlug
\resetcounters

\begin{document}

\title{A Population of Compact Radio Sources at the Galactic Center}
\author{Jun-Hui Zhao,$^1$ Mark R. Morris,$^2$ and W. M. Goss$^3$
\affil{$^1$CfA, Cambridge, MA 02138, USA; \email{jzhao@cfa.harvard.edu}}
\affil{$^2$UCLA, Los Angeles, CA 90095, USA; \email{morris@astro.ucla.edu}}
\affil{$^3$NRAO, Socorro, NM 87801, USA; \email{mgoss@nrao.edu}}}

% This section is for ADS Processing.  There must be one line per author.
\paperauthor{Jun-Hui Zhao}{jzhao@cfa.harvard.edu}{}{CfA | Harvard & Smithsonian}{SMA | RG Division, MS 78}{Cambridge}{MA}{02138}{USA}
\paperauthor{Mark R. Morris}{morris@astro.ucla.edu}{}{UCLA}{Department of Physics and Astronomy}{Los Angeles}{CA}{90095}{USA}
\paperauthor{W. M. Goss}{mgoss@nrao.edu}{}{NRAO}{P.O. Box O}{Socorro}{NM}{87801}{USA}

\begin{abstract}
The radio bright zone (RBZ) at the Galactic center has been observed with the JVLA in 
the A, B and C array configurations at 5.5 and 9 GHz. With a procedure
for high-dynamic range imaging developed on CASA, we constructed deep images  
at a resolution up to 0.2",
achieving rms noises of a few $\mu$Jy/beam. From the high-resolution and high-dynamics range images at 5.5 and 9 GHz, a population of compact radio sources ranging from a few mJy to
a few tens $\mu$Jy in flux density is revealed. 
The steep-spectrum radio sources in the RBZ are likely the candidates of high-energy objects that are associated with neutron stars and/or stellar
mass black holes at the Galactic center. 
We report new results of the Cannonball and Galactic center transient (GCT). 
\end{abstract}

\section{High Dynamic Imaging}
Utilizing the technique discussed in ZMG 2019, we constructed a high-dynamic image with
the JVLA data observed at 5.5 GHz in A, B and C arrays, by adding old VLA D-array data as well as
GBT single dish data. Fig. 1 shows the final image made with the hybrid data sets sampled in the UV
domain from 0 to 800k$\lambda$, achieving rms 2 $\mu$Jy/beam or a dynamic range 360,000:1.
We also imaged RBZ with only the large arrays' data. For 5.5 GHz, the image made using the A array data with Brigg's weight
(R=0) while for the 9 GHz image, both A and B arrays' data were used with R=0.5.
Both images are convolved to a common beam 0.44"x0.20" (1.8 deg), with compatible rms of 3 $\mu$Jy/
beam. The images are used to catalog the compact radio sources.

{\bf Cannonball} was discovered in hard X-ray (Park et al 2005). The radio counterpart with a peak
radio intensity of 500 $\mu$Jy/beam was detected with the VLA at both 5.5 and 8.3 GHz suggesting
presence of pulsar wind nebula (PWN) with relatively flat spectra for the head and plume ($\alpha$=-0.44
and -0.1) and a steep spectrum tail ($\alpha$=-1.94) (ZMG2013). From two epochs' data, the proper motion
of $\mu_\alpha = 0.001\pm0.003$ asec/y and $\mu_\delta = 0.013\pm0.003$ asec/y was inferred, 
suggesting 500 km/s in
transvers velocity. Adding two new epochs' data, we derived a consistent result with better statistics:
$\mu_\alpha = 0.0004\pm0.0025$ asec/y and  $\mu_\delta = 0.0130\pm0.0027$ asec/y, suggesting 
$V_{\rm t} = 483 \pm 94$ km/s. The age
of 9300 y for Sgr A East SNR is inferred.

{\bf GCT} was caught during its flare in 1990, 
reaching a peak in January 1991 with a flux density $\sim$0.9 Jy at 1.5 GHz greater than Sgr A*, 
and then subsequently declined at a rate of $\sim$22 mJy/day (Zhao et al. 1992). 
With the JVLA, we have detected significantly a compact
object in radio the intensity of 61$\pm$3 $\mu$Jy/beam at 9 GHz in 2015 and 91$\pm$10 
$\mu$Jy/beam at 5.5 GHz in 2014 at the position of the GCT. Assuming the GCT is 
in a quiescent state, no variations in flux density during 2014-2015
epochs, the spectral index of $\alpha = -0.81\pm0.25$ is inferred, 
consistent with $\alpha$ = -1.2 during the 1990 flare.

\articlefigure{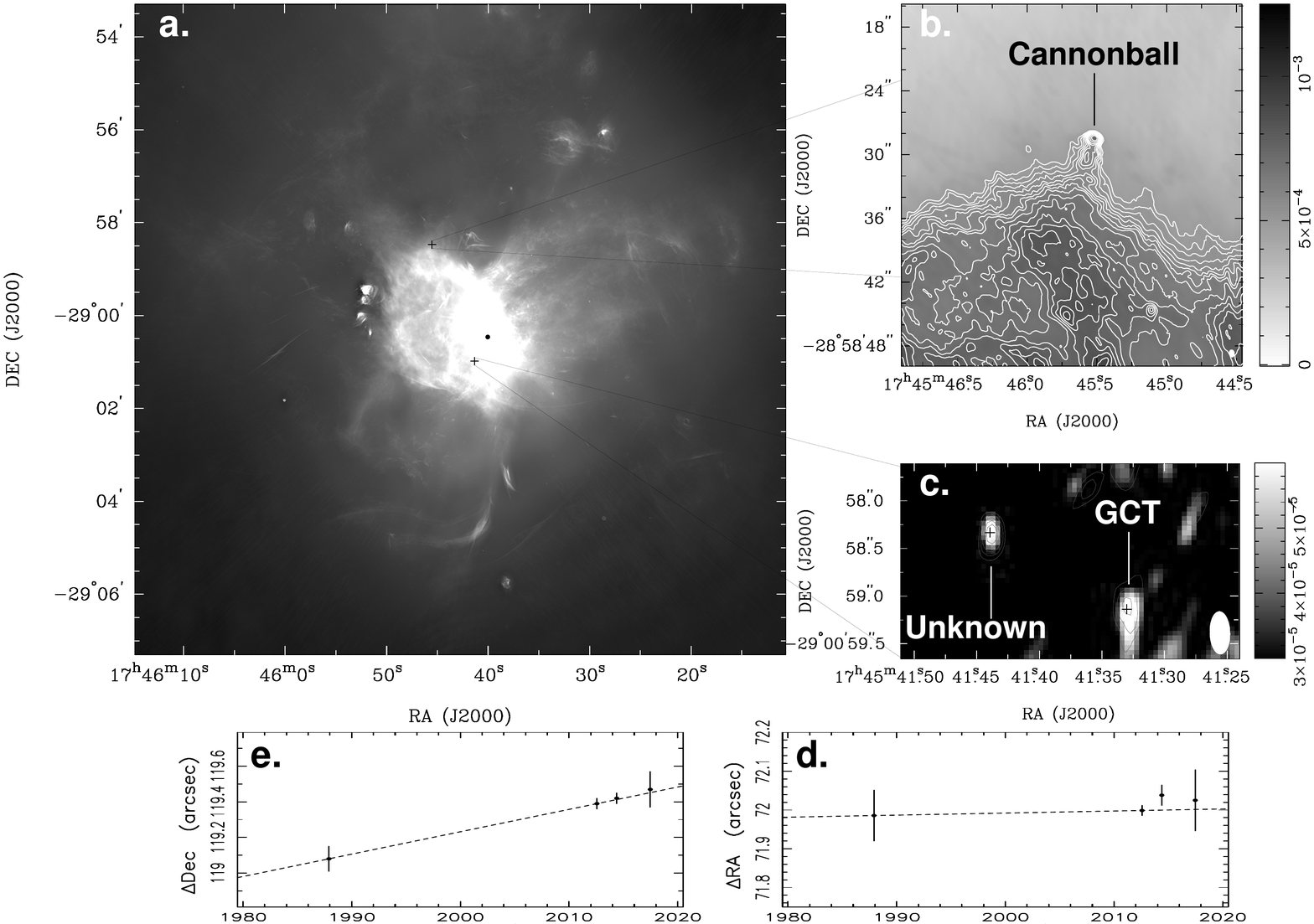}{ex_fig1}{{\bf a.} Image of RBZ at 5 GHz, with rms of 2 $\mu$Jy/beam and FWHM beam 0.68"x0.47" (8.4 deg).
{\bf b.} The Cannonball region. {\bf c.} The GCT region, greyscale for 5.5 GHz and contours for 9 GHz;
The rms noises are  $\sim$3 $\mu$Jy/beam, with FWHM beam of 0.44"x0.2" (1.8 deg). In addition to
the GCT, a previous unknown source with a flat spectrum $\alpha=-0.14$ is also detected.
{\bf d.} Proper motion fit to RA offset of the Cannonball from Sgr A*. {\bf e.} Proper motion fit to Dec 
offset.}

\section{Summary}
From our deep images of the RBZ at the Galactic center,
at least
a thousand of compact radio sources at a level of a few 100 $\mu$Jy in flux density
at 5 GHz appear to hide in the overwhelming diffuse continuum emission from
the region. High-resolution (0.2" images at 5.5 and 9 GHz have revealed the
presence of a sub-group of high energy objects, among the population of
compact radio sources, that are associated with non-thermal radiation. The
characteristics in radio spectrum suggest that they are the candidates related to
binary systems bounded with either neutron stars or stellar mass black holes.
% For non-BibTex:

\end{document}